\documentclass[sigconf]{acmart}

\usepackage{balance}

\AtBeginDocument{%
  \providecommand\BibTeX{{%
    \normalfont B\kern-0.5em{\scshape i\kern-0.25em b}\kern-0.8em\TeX}}}

\copyrightyear{2025}
\acmYear{2025}
\setcopyright{cc}
\setcctype{by}
\acmConference[SIGIR '25]{Proceedings of the 48th International ACM SIGIR Conference on Research and Development in Information Retrieval}{July 13--18, 2025}{Padua, Italy}
\acmBooktitle{Proceedings of the 48th International ACM SIGIR Conference on Research and Development in Information Retrieval (SIGIR '25), July 13--18, 2025, Padua, Italy}\acmDOI{10.1145/3726302.3730363}
\acmISBN{979-8-4007-1592-1/2025/07}

\begin{document}

\title{Second SIGIR Workshop on Simulations for Information Access (Sim4IA 2025)}
\author{Philipp Schaer}
\orcid{0000-0002-8817-4632}
\affiliation{%
  \institution{TH Köln}
  \city{Cologne}
  \country{Germany}
}
\email{philipp.schaer@th-koeln.de}

\author{Christin Katharina Kreutz}
\orcid{0000-0002-5075-7699}
\affiliation{
\institution{TH Mittelhessen}
\city{Gießen}
\country{Germany}
}
\email{ckreutz@acm.org}

\author{Krisztian Balog}
\orcid{0000-0003-2762-721X}
\affiliation{%
  \institution{University of Stavanger}
  \city{Stavanger}
  \country{Norway}}
\email{krisztian.balog@uis.no}

\author{Timo Breuer}
\orcid{0000-0002-1765-2449}
\affiliation{%
  \institution{TH Köln}
  \city{Cologne}
  \country{Germany}
}
\email{timobreuer@acm.org}

\author{Andreas Konstantin Kruff}
\orcid{0009-0002-8350-154X}
\affiliation{%
  \institution{TH Köln}
  \city{Cologne}
  \country{Germany}
}
\email{andreas.kruff@th-koeln.de}


\begin{abstract}
Simulations in information access (IA) have recently gained interest, as shown by various tutorials and workshops around that topic. 
Simulations can be key contributors to central IA research and evaluation questions, especially around interactive settings when real users are unavailable, or their participation is impossible due to ethical reasons. In addition, simulations in IA can help contribute to a better understanding of users, reduce complexity of evaluation experiments, and improve reproducibility. 
Building on recent developments in methods and toolkits, the second iteration of our Sim4IA workshop aims to again bring together researchers and practitioners to form an interactive and engaging forum for discussions on the future perspectives of the field. An additional aim is to plan an upcoming TREC/CLEF campaign. 
\end{abstract}




\maketitle

\section{Motivation}

Information access (IA) systems (like search engines, recommender systems, or conversational systems) are complex, and evaluating them is even more challenging. In domains like information retrieval, evaluation is closely coupled to the Cranfield paradigm, the dominant evaluation method in the field to deal with the inherent complexity in the information retrieval context. 
But Cranfield is not uncriticized \citep{ingwersen_turn_2005}. The underlying assumptions can lead to (over-)simplifications and potentially unrealistic search evaluations.
Therefore, other evaluation methods, including interactive retrieval settings \citep{kelly_methods_2009}, living labs \citep{jagerman-2017-overview}, or (user) simulation studies \citep{balog2023user} were proposed and discussed. 
These alternative evaluation endeavours aim to enable a more realistic and holistic information access evaluation by including richer user models or more complex representations of the search processes (like sessions). 

Simulations in information access have recently gained interest, as shown by various tutorials and workshops around that topic. 
A considerable number of relevant papers on user simulations were accepted, and even a study on simulating user queries won the best paper award at ECIR 2022 \citep{DBLP:conf/ecir/PenhaCH22}. Additionally, since the introduction of generative AI methods into the field, the possibility of integrating LLMs to simulate users has opened up a new chapter. 
Aside from the first iteration of our Sim4IA workshop at SIGIR 2024~\citep{DBLP:conf/sigir/SchaerKB0F24}, no track or workshop on simulations was present at ECIR or SIGIR in 2022--24. Additionally, no specific tracks or labs on simulations exist at TREC or CLEF. 

Simulations can be key contributors to central IA research and evaluation questions, especially around interactive settings when real users are unavailable, or their participation is impossible due to ethical reasons. In addition, simulations in IA can help contribute to a better understanding of users, reduce complexity of evaluation experiments, and improve reproducibility. 
Formalizing a user model for simulation delivers explicit hypotheses on user behavior, which can produce insights into the validity of assumptions on users \citep{balog2023user}. 

The Workshop on Simulations for Information Access (Sim4IA) at SIGIR 2024~\citep{DBLP:conf/sigir/SchaerKB0F24} was the latest example of the re-started interest in the topic of simulation. Around 25 participants discussed recent developments and challenges. The main conclusion from the workshop's discussion was that, while the interest is high, challenges with respect to how to model a shared task remain unsolved. Questions like evaluating the validity of simulated users within a shared task setting or what kinds of user archetypes are worth considering in this context are still open. The main takeaway was that there is a need for additional community work and that the participants envisioned a follow-up event to this workshop for having a more focused and in-depth discussion with experienced shared task organizers and participants~\citep{breuer2024reportworkshopsimulationsinformation}.


\section{Related Workshops}

At SIGIR 2024 we held the first iteration of our workshop~\citep{DBLP:conf/sigir/SchaerKB0F24} with around 25 participants~\citep{breuer2024reportworkshopsimulationsinformation}.
In addition to our workshop, two others directly related to IR simulations have been held so far, SIGINT \citep{DBLP:journals/sigir/AzzopardiJKS10} in 2010 and Sim4IR \citep{DBLP:journals/sigir/BalogMTZ21} in 2021. 
SIGINT focused on simulating interactions mainly for interactive IR. 
The potential of simulations as an evaluation methodology beyond static test collection experiments was explored. 
Sim4IR further explored the capabilities of simulation-based evaluations. 
Amongst other topics, it was discussed how accurate evaluations need to reflect human behavior, how simulated and human judges relate, and how simulations as an evaluation methodology interplay with established methods. 


\section{Workshop Goals}

In this current and second iteration of the workshop\footnote{\url{https://sim4ia.org/sigir2025/}}, we dive deeper into the field by focusing on the user simulation part. This form of simulation recently gained popularity due to available toolkits like 
SimIIR 3.0\footnote{\url{https://github.com/simint-ai/simiir-3}}~\cite{simiir3} or tutorials on this exact topic (CIKM 2023~\cite{DBLP:conf/cikm/BalogZ23}). 
Additionally, recent venues like CLEF Touché Lab~\citep{10.1007/978-3-031-56069-9_64} and TREC iKAT~\citep{aliannejadi2024trec} employed user simulations in conversational settings, demonstrating a growing interest in these matters, as well as possibilities of LLM-based approaches. 
Nevertheless, a specific venue to present and discuss new and experimental approaches and evaluation settings is missing. 

To understand how and whether the evaluation of information access technology can truly benefit from simulating user interactions, not only tools and frameworks are critical, but a multidisciplinary discussion and mutual understanding among the broad and sometimes conflicting perspectives is necessary. Simulations have to be aligned to the real-world settings of users and their complex information needs, contexts, and requirements. This workshop should serve as a forum to bring together researchers and practitioners. Additionally, this workshop aims to provide a much-needed forum for the community to discuss the emerging challenges when applying (user) simulations to evaluate information access systems in simulation-based shared tasks. Our goals are to:

\begin{itemize}
    \item Continue our series of workshops to generate an open conversation about possible future scenarios, applications, and methods to include simulations in the evaluation of IA systems;
    \item Provide a forum at SIGIR to discuss the pressing and emerging issues the IR community faces, and how simulations can help to overcome these;
    \item Develop and advertise the idea of organizing a TREC/CLEF campaign that includes simulations as a core element;
    \item Test an initial setting for two (micro) shared tasks designed around two IA use cases that might form the basis for the aforementioned TREC/CLEF campaign.
\end{itemize}

\section{Micro Shared Tasks}

\begin{figure}
    \centering
    \includegraphics[width=\linewidth]{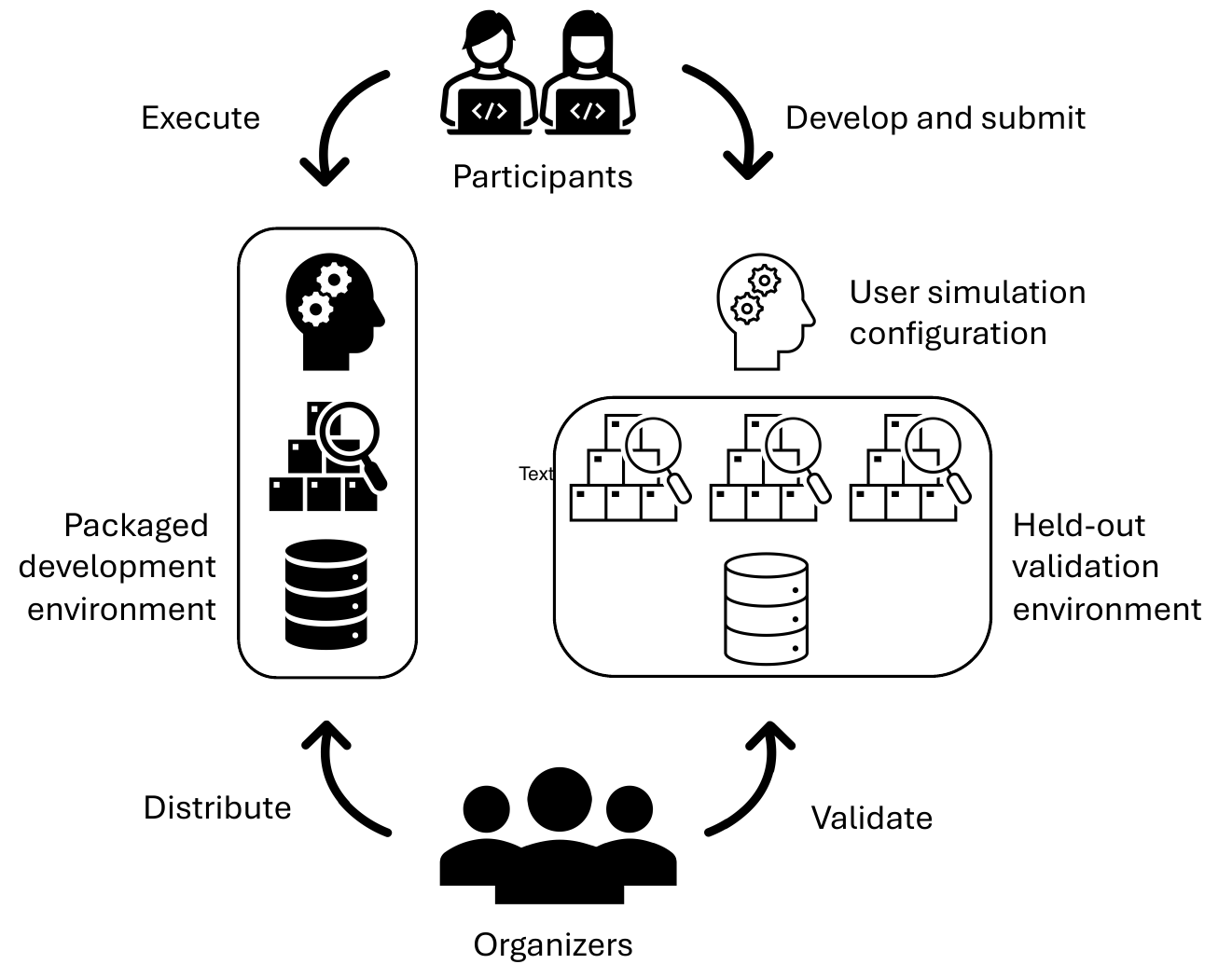}
    \caption{Shared task design that provides participants with prepared development environments that help defining user simulation configurations, which will be evaluated by the organizers with held-out data and systems.}
    \label{fig:shared_task}
\end{figure}


While running the Sim4IA workshop in 2024~\cite{breuer2024reportworkshopsimulationsinformation}, we encountered generally strong interest among the participants in running a shared task that builds on user simulations. Still, we could not flesh out the idea because there were too many uncertainties. With this year's workshop, we intend to define a shared task for TREC/CLEF on the basis of two micro shared tasks, which we will tackle in Sim4IA. Table~\ref{tab:timeline} gives a tentative timeline for our micro shared task.

\subsection{Concept}
Our anticipated shared task concept is based on the fundamental design principle of validating user simulations instead of measuring system effectiveness. 
That is, we envision users interacting with a particular IA system, such as a traditional search engine (Task A) or a conversational system (Task B), and challenge participants to design and implement user simulators that can mimic the interactions of real users with these systems with a high degree of fidelity. 
The workshop features a stripped-down version of this concept, which we deem micro shared tasks.



Figure ~\ref{fig:shared_task} illustrates the general setting where participants focus on developing user simulators. To lower the barrier to entry and allow participants to focus on the more (scientifically) interesting aspects of the task as opposed to system engineering, we provide them with data, baseline systems, and baseline user configurations packaged in a dockerized environment.
Specifically, we plan to employ the SimIIR simulation framework. Recently released as version~3, this framework provides the simulation of both conversational users and users within traditional ad-hoc retrieval systems, offering the advantage of operating within a single ecosystem~\cite{simiir3}. With its latest iteration, the framework now supports large language models for all existing user actions. The use of LLM-based prompting strategies is not only a timely topic, making it an intriguing choice for a workshop task, but it also offers a lightweight and straightforward introduction to the framework and the subject matter.

Thus, participants are given an information access system like PyTerrier as well as a simple baseline user simulator that can interact with that system for both tasks. 
Participants can then directly focus on developing user configurations and designing prompts without the overhead of setting up the simulation environment. 
Upon completion, participants submit their configurations, accompanied by a short system description explaining their design choices. 
The submitted user simulators are evaluated on the basis of held-out data collected from real users, using simple measures.

This setup aims to reduce complexity, leave room for extensions, and provide a shared ground for discussion. We note that questions around the evaluation of simulators (e.g., what measures to use) are meant to be an integral part of the workshop's discussion. Participants are invited to propose reasonable concepts for validating the simulations within the constraints of the given setting.
Ultimately, our goal is to leverage what we learn from best practices and potential pitfalls for the sake of a more fleshed-out concept that can be proposed to TREC/CLEF.




\subsection{Task A - Interactions with ranked lists} 
Task A focuses on simulating user interactions in a classic retrieval search session with ranked item lists. Participants conceptualize and implement valid user simulators for interactions with query-based retrieval. They develop solutions for query formulation in search sessions. 
Task A will be built on a newly crafted dataset from the LongEval 2025 lab~\citep{10.1007/978-3-031-88720-8_58} at CLEF\footnote{\url{https://clef-longeval.github.io/}} based on interaction data collected from the CORE\footnote{CORE (COnnecting REpositories): \url{https://core.ac.uk/}} platform. The data set will consist of three main components that contain search queries with their corresponding SERPs and click information, so that simulations can be based on previous queries and interactions with the result lists.

\subsection{Task B - Conversational systems} 

Task B focuses on simulating user utterances in a conversational search setting. Given a sequence of conversational utterances between a user and a system, where the last utterance was from the system, the task is to simulate the next user utterance. The conversational data used for this task is generated synthetically based on traditional search logs. Evaluation will be performed based on the semantic similarity of the predicted vs. actual user utterances.

\section{Format and Structure}

\begin{table}[t]
    \caption{Timeline for micro shared tasks}
    \label{tab:timeline}
    \vspace{-0.5\baselineskip}
    \centering
    \begin{tabular}{lp{6.5cm}}
        \toprule
        Time & Task \\
        \midrule        
        February  & Promotion of workshop and micro shared tasks via social media platforms (e.g., X, LinkedIn), distribution of call via mailing lists (e.g., SIG-IRList)\\
        16th May & Start of micro shared task and release of the training dataset and Docker SimIIR package\\
        13rd June & Release of the test dataset \\
        27th June & Submission deadline for simulation configurations\\
        4th July & Submission deadline for Lab Notes \\
        17th July & Workshop at SIGIR in Padua, Italy\\
        \bottomrule
    \end{tabular}
\end{table}
\begin{table}[t]
    \caption{Tentative schedule for the workshop.}
    \label{tab:schedule}
    \vspace{-0.5\baselineskip}
    \centering
    \begin{tabular}{ll}
    \toprule
    Time & Agenda\\ \midrule
        9:00--9:15 & Welcome \\
        9:15--10:00 & Keynote \\
        10:00--10:30& Invited tech talks on toolkits and infrastructure\\
        11:00--12:00 & Panel discussion\\
        12:00--12:30 & Overview talk on shared tasks\\
        13:30--14:15 & Lightning talks on shared tasks\\
        14:15--15:00 & Breakout group discussions pt 1\\ 
        15:30--16:30 & Breakout group discussions pt 2\\
        16:30--17:00 & Reports of the group discussions and closing \\
       \bottomrule
    \end{tabular}
\end{table}

We anticipate a \textit{highly interactive} and \textit{engaging}, \textit{full-day} workshop to foster collaborations and bring the community together. 
We are keen to hear different views and opinions on the current and future directions for simulations in information access systems. 
Therefore, we plan a keynote, panel discussion, tech talks, lightning talks on our micro shared tasks, and guided breakout sessions.
Table~\ref{tab:schedule} gives a tentative schedule. The morning sessions will be directed towards a broader audience with potentially less actual previous experience in simulation, while the afternoon sessions will focus more on the participants of the micro shared tasks.




\textbf{Keynote.} 
We strive to invite a recognized researcher representing industry and academia viewpoints from the field.

\textbf{Tech talks and lightning talks.}
We allocate short slots (approx. 10 minute talks) for invited tech talks on existing simulation toolkits and infrastructure components and accepted working papers to the two shared tasks for presentation.

\textbf{Panel discussion.}
We will moderate a panel with invited panellists from a diverse set of backgrounds with different levels of seniority championing different sub-fields and views. As we plan for an upcoming TREC/CLEF track/lab, we would like to invite someone from those communities/steering committees as well. Our keynote speaker will be joined by four more panellists.

\textbf{Overview talk on shared tasks.}
We will give a short TREC-style overview talk on common problems or strategies encountered in the two micro shared tasks.

\textbf{Breakout group discussions.}
Attendees of the workshop will be split into smaller groups, each focusing on different topics or questions. 
One task for the group work will be the conceptualization of a TREC/CLEF campaign based on the learnings from the two micro shared tasks.
Other topics could include the specification of personas for evaluations which could become a guideline to refer back to or
the extension of pre-existing user models with new components and their incorporation into evaluation measures. 
Breakout groups' discussions will be moderated by some previously generated questions.
Ideas, opinions, and outcomes of the groups are captured in shared Google Docs.

\textbf{Reports of the group discussions and closing.}
Joint discussion and merging of results from the breakout group discussions. 
Final notes and discussion of future directions for the community.

\section{Contributions and Outcome}

Our workshop is expected to attract a diverse audience of researchers and practitioners from the field of IR, and we expect to have at least 25 participants.
We invite contributions in the form of code, prompts, or simulators targeting one or both of our two micro shared tasks, as well as a brief lab note motivating the prompt design, applied methods, encountered problems, and open questions. We do not expect long and detailed papers. Therefore, the working notes will be editorially reviewed by the organizers and later shared together with the simulation configurations on Zenodo. 

Our goal is to strengthen the community, connect researchers at different academic stages, and foster a common understanding of fundamental concepts in the topic of the workshop.
The conceptualization of a TREC/CLEF campaign will be followed up on by the organizers and other interested participants.
Authors of accepted working notes for the micro shared tasks will be invited to join as co-authors on a SIGIR Forum paper.

\section{Organizers}
The following organizers will contribute to this workshop.

\textbf{Philipp Schaer} holds a focus professorship for data science (previously information retrieval) at the Institute of Information Science (TH Köln -- University of Applied Sciences, Cologne, Germany). He published at IR-related venues like ECIR, SIGIR, and JCDL on topics like evaluation, digital libraries, and reproducibility. Recently, he was co-program chair of ACM/IEEE JCDL 2022, co-organizer of the CLEF LiLAS lab and Sim4IA at SIGIR'24. 

\textbf{Christin Kreutz} is a tandem professor for data science in the humanities working in industry and academia. Her research focuses on user aspects in IA systems and digital libraries. She has been co-organizing the Sim4IA'24 and SCOLIA'25 workshops.

\textbf{Krisztian Balog} is a full professor at the University of Stavanger and a staff research scientist at Google DeepMind. 
His research focuses on intelligent information access, with a current emphasis on conversational systems and user simulation.
Balog has co-organized numerous workshops at SIGIR and CIKM (including Sim4IR at SIGIR'21 and Sim4IA at SIGIR'24) as well as large-scale benchmarking efforts at TREC and CLEF, and has recently co-authored an FnTIR monograph on user simulation~\citep{balog2023user}.

\textbf{Timo Breuer} is a postdoctoral researcher interested in reproducible IR evaluations and user-oriented living lab experiments. He sees user simulations as a key element to bridge the gap between these two disciplines.

\textbf{Andreas Kruff} is a PhD student at TH Köln, where he is working on issues related to the reproducibility of interactive IR evaluation settings. His special focus is simulation studies and how they can help to improve reproducibility.

\begin{acks}
This work is partially funded by Deutsche Forschungsgemeinschaft (DFG) under grant number 509543643 and within the funding programme FH-Personal (PLan CV, reference number 03FHP109) by the German Federal Ministry of Education and Research (BMBF) and Joint Science Conference (GWK).

\end{acks}

\bibliographystyle{ACM-Reference-Format}
\balance
\bibliography{main}

\end{document}